\documentclass[assymb,11pt]{article}
\usepackage[pdftex,colorlinks=true,urlcolor=black,filecolor=black,linkcolor=black,
            pdftitle={Einstein scalar wave.},
            pdfauthor={Mark D. Roberts},
            pdfsubject={relativity,exact solutions},
            pdfkeywords={wave,scalar,exact solutions},
            pagebackref,pdfpagemode=None,bookmarksopen=true]{hyperref}
\usepackage{amssymb}
\newcommand{\bc}{\begin{center}}
\newcommand{\ec}{\end{center}}
\newcommand{\bi}{\begin{itemize}}     
\newcommand{\ei}{\end{itemize}}
\newcommand{\bd}{\begin{description}} 
\newcommand{\ed}{\end{description}}
\newcommand{\bn}{\begin{enumerate}}   
\newcommand{\en}{\end{enumerate}}
\newcommand{\be}{\begin{equation}}
\newcommand{\ee}{\end{equation}}
\newcommand{\ber}{\begin{eqnarray}}
\newcommand{\ear}{\end{eqnarray}}
\newcommand{\ba}{\begin{array}}
\newcommand{\ea}{\end{array}}
\newcommand{\bv}{\begin{verbatim}}
\newcommand{\ev}{\end{verbatim}}

\newcommand{\n}{\nonumber\\}

\begin{document}
\title{Causality in scalar-Einstein waves.}
\author{
\href{http://www.violinist.com/directory/bio.cfm?member=robemark}
{Mark D. Roberts},\\
Flat 44,  The Cloisters,  83 London Road,  Guildford,  GU1 1FY,  UK\\
}
\date{$10^{th}$ of March 2015}
\maketitle
\begin{abstract}
A wavelike scalar-Einstein solution is found and indicating vectors constructed from
the Bel-Robinson tensor are used to study which objects co-move with the wave and whether
gravitational energy transfer is null.
\end{abstract}
\tableofcontents
\section{Introduction}
Gravitational waves are usually taken to move at the speed of light.
Among the many questions that arise are whether the waves gravitational energy travels at
the same speed and whether it is possible to have co-moving fields:
in particular what happens when the wave moves from a vacuum to a medium such as dust,
in this case do both gravity and co-moving fields slow down by the same amount.
To this end a new solution which contains both gravitational wave and has a scalar field
obeying
\be
Rs_{ab}\equiv R_{ab}-2\phi_a\phi_b=0,
\label{rscalar}
\ee
is found and investigated by constructing indicator vectors
produced by transvecting the Bel-Robinson tensor.
\section{The plane wave.}\label{planew}
\subsection{The plane wave line element}\label{pple}
The plane wave has line element \cite{HE}
\be
ds^2=W(u,y,z)du^2+2dudv+dy^2+dz^2,
\label{planele}
\ee
the determinant of the metric is $g=-1$ and the Kretschmann curvature invariant vanishes $K=RiemSq=0$,
the non-vanishing components of the Riemann tensor are given by
\be
R_{uiuj}=-\frac{1}{2}W_{,ij}
\label{pwrie}
\ee
where $i,j\dots=1,2,3$.  The non-vanishing component of the Ricci tenor is
\be
R_{uu}=-\frac{1}{2}W_{,yy}-\frac{1}{2}W_{,zz}.
\label{pwric}
\ee
\subsection{The vacuum,  Bach and scalar plane wave}\label{ppv}
For a vacuum Ricci flat spacetime a choice of $W$ is
\be
W=(y^2-z^2)f(u)-2yzg(u)
\label{pww}
\ee
where $f,g$ are arbitrary twice differentiable functions of $u$.
The Bach tensor is
\be
B_{ab}\equiv 2C_{a..b}^{~cd}R_{cd}+4C_{a..b;cd}^{~cd},
\label{defbach}
\ee
and this tensor can be used in the expression for quadrtaic field equations
\be
R_{ab}+bB_{ab}=0.
\label{bfq}
\ee
For the line element (\ref{planele}) the Bach tensor has non-vanishing component
\be
B_{uu}=W_{,yyyy}+2W_{,yyzz}+W_{,zzzz}.
\label{bpw}
\ee
A solution to the field equations (\ref{bfq}) is
\be
W=\sin\left(\frac{y}{\sqrt{b}}\right)f_1(u)
+\cos\left(\frac{y}{\sqrt{b}}\right)f_2(u)
+\sin\left(\frac{z}{\sqrt{b}}\right)g_1(u)
+\cos\left(\frac{z}{\sqrt{b}}\right)g_2(u).
\label{wq}
\ee
For a scalar-Einstein solution one can choose
\be
W=(ay^2-bz^2)f(u)-2cyzg(u),
\label{abW}
\ee
giving
\be
R_{uu}=(b-a)f=2\phi_u^2,
\label{rsc}
\ee
however $\phi$ can have no $y,z$ dependence as this would entail non-vanishing $R_{yy}$
and in this sense the scalar field is not co-moving with the gravitational field.
\subsection{Null tetrads for the plane wave}\label{ntpp}
A suitable set of null tetrads is
\begin{equation}
l_a=-\frac{1}{2}W\delta^u_a-\delta^v_a,~~~
n_a=\delta^u_a,~~~
m_a=-\frac{i}{\sqrt{2}}\delta^y_a-\frac{1}{\sqrt{2}}\delta^z_a,
\label{lnmpp}
\end{equation}
the Weyl and Ricci scalars are
\begin{equation}
\Phi_{22}=-\frac{3}{4}\left(W_{yy}+W_{zz}\right),~~~
\Psi_4=\frac{3}{4}\left(W_{yy}+2iW_{yz}-W_{zz}\right),
\label{wrpp}
\end{equation}
for the particular case (\ref{abW}) with $a,b=1$ (\ref{wrpp}) reduces to
\begin{equation}
\Phi_{22}=0,~~~
\Psi_4=3\left(f-ig\right).
\label{wrr}
\end{equation}
\section{A Scalar-Einstein wave.}\label{sew}
\subsection{The line element}\label{sele}
For a $y,z$ dependent scalar-Einstein wave consider the line element
\be
ds^2=W(u,x,y)du^2+2A_3xydudv+A_1dx^2+A_2dy^2,~
\phi=\frac{1}{2}\ln\left(\frac{kx}{y}\right),
\label{sewave}
\ee
in the case of vanishing gravitational wave it is related to the solution in \cite{mdr13},
although the corresponding conformal Killing vector has not been found.
After subtracting off the scalar field (\ref{rscalar}) there remains the Ricci tensor component
\be
Rs_{uu}=-\frac{1}{2A_2y^2}\left(W-yW_{,y}+y^2W_{,yy}\right)
       -\frac{1}{2A_1x^2}\left(W-xW_{,x}+x^2W_{,xx}\right).
\label{riccisew}
\ee
The line element (\ref{sewave}) is a scalar-Einstein solution when
\ber
W&=&\left(B_1x{\rm BesselJ}\left(0,\frac{x}{\sqrt{A_2}}\right)+B_2x{\rm BesselY}\left(0,\frac{x}{\sqrt{A_2}}\right)\right)\\
&\times&\left(C_1y{\rm BesselJ}\left(0,\frac{y}{\sqrt{-A_1}}\right)+C_2y{\rm BesselY}\left(0,\frac{y}{\sqrt{-A_1}}\right)\right)f(u),
\nonumber
\label{bessel}
\ear
lowest order expansion suggests that the $C_1$ term is real but the $C_2$ term might be complex.
\subsection{A Simpler Case}\label{simpc}
A more simple solution to (\ref{riccisew}) is
\be
H=\left(B_1x+B_2x\ln(x)\right)\left(C_1y+C_2y\ln(y)\right)f(u),
\label{vacsew}
\ee
the invariants can be expressed in terms of the Ricci scalar
\be
R=\frac{A_1x^2+A_2y^2}{2A_1A_2x^2y^2},
\label{ricciscalar}
\ee
and are
\ber
\label{inv}
&&K=3R^2,~
WeylSq=\frac{4}{3}R^2,~
RicciSq=R^2,~
BS=\frac{4}{9}R^4\\
&&R_1=\frac{3}{16}R^2,~
R_2=\frac{3}{64}R^3,~
R_3=\frac{21}{1024}R^4,
W_{1R}=\frac{1}{6}R^2,~
W_{2R}=\frac{1}{36}R^3,\n
&&M_{2R}=M_3=\frac{1}{96}R^4,~
M_{4}=\frac{1}{768}R^5,~
M_{5R}=\frac{1}{576}R^5.
\nonumber
\ear
\subsection{Null tetrads for the scalar-Einstein wave}\label{ntse}
A suitable set of null tetrads is
\begin{eqnarray}
&&l_a=-\frac{1}{2}xyH(u))(B_1+B_2\ln(x))(C_1+C_2\ln(y))\delta^u_a-A_3xy\delta^v_a,\n
&&n_a=\delta^u_a,~~~
m_a=\frac{A1}{\sqrt{_2}}\delta^x_a+\frac{iA_2}{\sqrt{2}}\delta^y_a,
\label{lnmse}
\end{eqnarray}
the Weyl and Ricci scalars are
\begin{eqnarray}
&&\Phi_{02}=\frac{(\sqrt{A_1}x-\sqrt{A_2}y)^2}{8A_1A_2x^2y^2},~~~
\Phi_{11}=\frac{1}{8}R,\n
&&\Psi_2=\frac{1}{6}R,~~~
\Psi_4=i\times \frac{poly H(u)}{4\sqrt{A_1A_2}A_3^2x^2y^2},\n
&&poly\equiv B_1C_2+B_2C_1+B_2C_2\left(2+ln(x)+\ln(y)\right).
\label{wres}
\end{eqnarray}
\section{The Bel-Robinson Tensor}\label{brt}
\subsection{Definition of the Bel-Robinson tensor}\label{dbr}
The dual of a tensor is defined by
\begin{equation}
*T_{abm...}=\frac{1}{2}\epsilon_{ab..}^{~~cd}T_{cdm...},
\label{ddual}
\end{equation}
where $m...$ are a set of indices.
The Bel-Robinson tensor is defined by
\begin{equation}
B_{cdef}\equiv C_{acdb}C_{.ef.}^{a~~b}+*C_{acdb}*C_{.ef.}^{a~~b},
\label{bellrob}
\end{equation}
The four-vector indicator of energy-momentum is
\begin{equation}
P_a=B_{abcd}V^bV^cV^d,
\label{4m}
\end{equation}
usually $V$ is taken to be a time-like vector field,  but other choices are possible:
for example when $V$ is replaced by the null tetrad vector $l$ the indicator
is referred to a $N_a$.
\subsection{The Bel-Robinson tensor for the Schwarzschild solution}\label{brsch}
The Schwarzschild solution has line element
\begin{equation}
ds^2=-\left(1-\frac{2m}{r}\right)dt^2+\frac{1}{1-\frac{2m}{r}}dr^2+r^2d\Sigma^2_2,~~~
d\Sigma^2_2\equiv d\theta^2+\sin(\theta)^2d\phi^2,
\label{sch}
\end{equation}
it has Weyl scalar
\begin{equation}
\Psi_2=-\frac{m}{r^3}.
\label{schw}
\end{equation}
For the time-like vector field
\begin{equation}
T_a\equiv f\delta^t_a,~~~
T_aT^a=-\frac{f^2}{1-\frac{2m}{r}},
\label{tsch}
\end{equation}
and the indicating four-vector (\ref{4m}) is
\begin{equation}
P_a=-\frac{6f^2\Psi_2^2}{1-\frac{2m}{r}}T_a,
\label{isch}
\end{equation}
which is conserved $P^a_{.;a}=0$.
\subsection{The Bel-Robinson tensor for the plane wave}\label{brpp}
The Bel-Robinson tensor for the plane wave has one component
\begin{equation}
B=B_{uuuu}=\frac{1}{4}\left(W_{zz}-W_{yy}\right)+W_{yz}^2,
\label{brp}
\end{equation}
the indicating four-vector is best (\ref{4m}) with $V$ replaced by $l$ from (\ref{lnmpp}) giving
\begin{equation}
N_a=Bn_a,
\label{ipp}
\end{equation}
which is conserved.
For the particular choice (\ref{abW}) with $a,b=1$
\begin{equation}
B=4\left(f^2+g^2\right).
\label{partpp}
\end{equation}
\subsection{The Bel-Robinson tensor for imploding scalar spacetime}\label{bres}
The line element is taken to be \cite{mdr13}
\begin{equation}
ds^2=-(1+2\sigma)dv^2+2dvdr+r(r-2\sigma v)d\Sigma_2^2,~~~
\phi=\frac{1}{2}\ln\left(1-\frac{2\sigma v}{r}\right).
\label{es}
\end{equation}
with $d\Sigma_2^2$ given by (\ref{sch}) and complementary null coordinate
$u\equiv (1+2\sigma)v-2r$.
In this case study of the Bel-Robinson tensor is not so straightforward and here is approached
by four methods.
\paragraph{Time-like vector method}\label{tlvm}
A time-like vector is
\begin{equation}
T_a=-\left(\frac{1}{2}(1+T_s)+\sigma\right)\delta^v_a+\delta^r_a,~~~
T_aT^a=-T_s,
\label{tles}
\end{equation}
and the indicating four-vector (\ref{4m}) is
\begin{equation}
P_a=-\frac{1}{6}T_sR^2T_a,
\label{pes}
\end{equation}
where $R$ is the Ricci scalar,  (\ref{pes}) is not conserved in general
\begin{equation}
P^a_{.;a}=\frac{2T_s\sigma^4uv}{3r^5(-r+2\sigma v)^5}\left(ut_1-T_svt_2\right),
\label{npes}
\end{equation}
where
\begin{equation}
t_1\equiv 3\sigma(1+2\sigma)v^2-(3+4\sigma)rv+2r^2,~~~
t_2\equiv t_1+2r(r-2\sigma v),
\label{teas}
\end{equation}
however it is conserved in the two particular cases $T_s=ut_1/vt_2$ and $T_s=0$
which is null and we go to next.
\paragraph{Null tetrad method}\label{npte}
The indicating four-vector (\ref{4m}) with $l$ replacing $V$ just gives $N_a=0$,
using mixtures of null tetrad vectors instead of $l$ no simple pattern arises.
\paragraph{Killing vector method}\label{kvm}
The solution (\ref{es}) has a homeothetic Killing potential $K=cuv$ wich can be partially
differentiated to give a homeothetic Killing vector,  using this in (\ref{4m}) gives
\begin{equation}
A_a=-\frac{2}{3}cKK_a,~~~
A^a_{.;a}=\frac{8}{3}c^2KR^2.
\label{kvi}
\end{equation}
\paragraph{Scalar field everywhere method}\label{sses}
In (\ref{4m}) one uses the gradient of the scalar field (\ref{es}) everywhere,  then
\begin{equation}
A_a=\frac{1}{12}R^3\phi_a,~~~
A^a_{.;a}=\frac{(r^2-(1+2\sigma)v(r-\sigma v))R^4}{2\sigma uv}.
\label{phiee}
\end{equation}
\subsection{The Bel-Robinson tensor for the scalar-Einstein wave}\label{brewave}
If one uses a time-like vector in (\ref{4m}) the no simple pattern arises,  however
for the null tetrads (\ref{lnmse}) one gets
\begin{equation}
N_a=-4A_3^4x^4y^4\Psi_4^2n_a.~~~
N^a_{.;a}=0,
\label{newave}
\end{equation}
The scalar field everywhere method gives
\begin{equation}
A_a=\frac{1}{12}R^3\phi_a,~~~
A^a_{.;a}=\frac{1}{2}R^4\frac{A_1x^2-A_2y^2}{A_1x^2+A_2y^2}.
\label{ewss}
\end{equation}
The stress of the scalar field gives scalar field propagation
\begin{equation}
P^a_\phi=T^{ab}V_b=-\frac{1}{2}RV^a,
\label{smstress}
\end{equation}
with similar equations for the null tetrad.
\section{Conclusion.}\label{conc}
The indicating four vector (\ref{isch}) is what one would want for the Schwarzschild solution,
it is timelike and furthermore co-directional with the chosen vector field (\ref{tsch});
however why the proportional function takes the form it does
does not seem to be predictable beforehand,
one might hope to identify $m$ as the overall energy just from the proportional function.
Similarly,  the indicating four vector (\ref{ipp}) is what one would want for the plane wave,
it is null and furthermore co-directional with the chosen null tetrad (\ref{lnmpp}).

The energetics of the imploding scalar solution (\ref{es}) are important:
the solution has no overall energy the negative energy of the gravitational field
and the positive energy of the scalar field cancel out,  and there is the question of
what happens locally where the scalar field energy can be measured but not the gravitational
field energy.
For the imploding scalar solution (\ref{es}) the null tetrad method 
gives vanishing indicator and the other three methods 
usually have non-vanishing conservation equation,  the exception being when $T_s=ut_1/vt_2$
in (\ref{npes}).
The non-vanishing of the conservation equations are of high order in $\sigma$.
This leads to the conclusion that there is no local balance in energy exchange:
it is only the global energy that cancels,
and any detail of how this can happen remains obscure.

For the scalar-Einstein wave 
the null indicator (\ref{newave}) is conserved,
suggesting that the square of gravitational energy is co-moving with the wave.
However the solutions scalar field does not seem to co-move (\ref{smstress}),
suggesting that the scalar field is an ambient medium.

A convenient property of the indicating four vector (\ref{4m}) is that it always turns out to
be proportional to a known vector of the spacetime,  for the non-null case this is the
transvecting vector,  for the null tetrad it is the complimentary null vector,  for example
the indicating four vector (\ref{4m}) is proportional to $n$ after transvecting with $l$:
this appears to be co-incidence.   That the conserved indicating vector for the scalar-Einstein
wave is null (\ref{newave}) indicates that the transfer of the square of its gravitational
energy is null and that the presence of the scalar field does not impede it thus answering the
question implied by the title.

\end{document}